\documentclass[preprint,letterpaper,sort&compress,12pt]{elsarticle}

\makeatletter
\def\ps@pprintTitle{%
  \let\@oddhead\@empty
  \let\@evenhead\@empty
  \let\@oddfoot\@empty
  \let\@evenfoot\@oddfoot
}
\makeatother
 
\usepackage{hyperref}
\hypersetup{
    colorlinks,
    citecolor=blue,
    filecolor=blue,
	    linkcolor=blue,
    urlcolor=blue
}

\newcommand {\beg}{\begin{equation}}
\newcommand {\en}{\end{equation}}

\newcommand {\begar}{\begin{array}}
\newcommand {\enar}{\end{array}}

\newcommand{\lef}[2]{\stackrel{\leftarrow#2}{#1}}
\newcommand{\rig}[2]{\stackrel{#2\rightarrow}{#1}}
\newcommand{\al}{\alpha}
\newcommand{\be}{\beta}

\begin{document}

\title{Integrable mappings for noncommuting objects}

\vspace{1cm}

\author{Andrei N. Leznov}
\address{Institute for High Energy Physics,142284 Protvino,
Moscow Region, Russia}

 \author{Emil A. Yuzbashyan\footnote{Email: eyuzbash@physics.rutgers.edu}}
 \address{Bogoliubov Laboratory of Theoretical Physics, Joint Institute for
Nuclear Research,
141980 Dubna, Moscow Region, Russia}

\begin{abstract}
We construct hierarchies of integrable systems invariant under the
two-dimensional Darboux-Toda mapping for noncommuting objects, thus generalizing to the noncommutative case the  integrable mapping approach to nonlinear dynamical systems.   Besides the usual setup with one time and two space dimensions, we consider the case when the unknown functions   also   depend on two
Grassman variables.
 \end{abstract}
 
 \maketitle 

\section{Introduction}
 
Recently, it has been suggested  \cite{1, 1_1}
that  the theory of integrable nonlinear systems is intimately  related to the representation theory of the group of integrable mappings (transformations of dynamical variables that preserve the form of equations of motion). However, the discussion so far has been limited to classical dynamical variables (classical fields), i.e. scalar functions of time and space coordinates. The purpose of the present paper is to  extend these ideas to  noncommutative settings, such as
  nonequilibrium quantum problems.

Presently, no classification of  integrable mappings is available even in the commutative case.
For this reason, we restrict ourselves with two specific examples of integrable mappings for noncommuting objects: Darboux-Toda transformation in the usual two-dimensional space and  in the two-dimensional superspace. For both mappings we explicitly
construct hierarchies of (1+2)-dimensional integrable systems of integro-differential equations of evolution type that are invariant under them. We will see that   our discrete symmetry (integrable mapping) based approach~\cite{1, 1_1}   is uniquely suited for
 deriving new (1+2)-dimensional integrable nonlinear systems.

The paper is organized as follows. In Section~\ref{DT_sec}, we construct the hierarchy of (1+2)-dimensional integrable systems invariant under the two-dimensional Darboux-Toda mapping for noncommuting objects. Subsections~\ref{DT_sec1},  \ref{DT_sec2}, and \ref{DT_sec3}
contain  definitions, main results, and concrete examples, respectively.  In Section~\ref{super_sec}, we consider the case when unknowns,  in addition to one time and two space coordinates, also depend on two Grassman variables. The  structure of Section~\ref{super_sec} is similar to that of Section~\ref{DT_sec}.

\section{Darboux-Toda mapping in two-dimensional space}
\label{DT_sec}

In this section, we construct an infinite set of integrable equations of motion for noncommuting objects  invariant under the two-dimensional Darboux-Toda mapping. 
 
\subsection{Definitions and the formulation of the problem}
\label{DT_sec1}

 We start by formulating the problem and introducing  notations that will be used throughout  the rest of the paper.
Let  $u$ and $v$ be a  pair of noncommuting functions of coordinates $t$, $x$, and $y$.
In particular, $u$ and $v$ can be square matrices, operators defined in a certain representation space, or
elements of a group.  Both $u$ and $v$ are assumed to be differentiable and invertible. No commutation relations are assumed. 

Darboux-Toda transformation is the following mapping of a pair of operators $(u, v)$
to a new pair $(\lef{u}{}, \lef{v}{})$ :
  \begin{equation}
  \begin{array}{cc}
\lef{u}{}= v^{-1},&\lef{v}{}= [vu-(v_x v^{-1})_y]v\equiv v[uv- (v^{-1} v_y)_x].
\end{array}
  \label{10}
  \end{equation}
{\emergencystretch=5pt
Transformation (\ref{10})  for square matrices   was  studied in \cite{10} in a different context. 
Mapping  (\ref{10}) is invertible, i.e. $u$ and $v$ can be explicitly expressed trough $\lef{u}{}$ and $\lef{v}{}$.  We write the inverse mapping as
\begin{equation} \begin{array}{cc}
  \rig{v}{}= u^{-1},&\rig{u}{}= [uv-(u_y u^{-1})_x]u\equiv u[vu- (u^{-1}u_y)_x].
  \end{array} \label{11}
  \end{equation}
We will also need a notation for a repeated application of mapping (\ref{10}) and its inverse
(\ref{11}) to functions of $u$, $v$, and their derivatives. Specifically,  the results of $s$ applications of direct and inverse transformations (\ref{10}) and (\ref{11})  to a function  $f(u,v)$  is denoted by $ \lef{f}{s}\equiv f(\lef{u}{s},\lef{v}{s})$
and
$\rig{f}{s}\equiv f(\rig{u}{s}, \rig{v}{s})$, respectively,
with a convention $\lef{f}{(-m)}\equiv\rig{f}{m}$ for $m\ge0$.
 }

 Now consider general evolution-type equations  for $u$ and $v$
\begin{equation}
\begin{array}{c}
u_t=F_1(u,v,u_x,v_x,u_y,v_y,...),\\
\\
v_t=F_2(u,v,u_x,v_x,u_y,v_y,...).\\
\end{array}\label{611}
\end{equation}
The system of equations~(\ref{611}) is said to be invariant under a given change of variables $u$ and $v$ if this change  preserves the exact form of  Eqs.~(\ref{611}). 
In particular, the system (\ref{611}) is invariant under the mapping (\ref{10}) if and only if the following
functional relations hold:
 \begin{equation}
\begin{array}{l} \lef{F}{}_1=
\lef{u}{}_t=-v^{-1}v_t v^{-1}=-v^{-1}F_2 v^{-1},\\ 
\\
\lef{F}{}_2=
\lef{v}{}_t=([vu-(v_x v^{-1})_y]v)_t=\\
\quad =[F_2 u+v F_1-(F_{2x}v^{-1})_y+
(v_xv^{-1}F_2v^{-1})_y]v+[vu-(v_xv^{-1})_y]F_2.\\
\end{array}   
\label{13}
\end{equation}
We call Eqs.~(\ref{13}) the symmetry equations for the mapping (\ref{10}). They  are
to be solved for   $F_1$ and $ F_2$ as \textsl{functions} of $u$, $v$, and their derivatives with respect to space coordinates $x$ and $y$. Note that the symmetry equations are linear with respect to $F_1$ and $F_2$ and that there is always a trivial solution $F_1=au_x+bu_y$ and $F_2=a v_x+
b v_y$, where $a$ and $b$ are arbitrary numbers.

\subsection{Solution of  the symmetry equations}
\label{DT_sec2}

In this subsection, we obtain an infinite set of solutions of the symmetry equations (\ref{13})
 by  properly  generalizing the approach of~\cite{14}.  

We start by rewriting the symmetry equation (\ref{13}) in a more convenient form.
Let 
\begin{equation}
F_2=\al_0 v, \qquad F_1=u\be_0.
\label{f}
\end{equation}
 In terms of variables $\al_0$ and $\be_0$ equations
(\ref{13}) read
\begin{equation}
\begin{array}{l}
\be_0=-\rig{\al}{}_0,\\ 
{\al_0}_{xy}=(\al_0-\lef{\al}{}_0)\lef{T_0}{\:}+
T_0(\al_0-\rig{\al_0}{\!})+\theta {\al_0}_y
-{\al_0}_y\theta, 
\label{2}
\end{array}
\end{equation}
where $T_0=vu$ and $\theta=v_x v^{-1}$.
These equations have an obvious solution ${\al_0}^{(0)}=-{\be_0}^{(0)}=1$ (note that
$\lef{1}{}=\rig{1}{}=1$).  From this solution, using
equations (\ref{f}) and (\ref{611}), we immediately   obtain the first term of the  hierarchy,
$$
u_t=-u,\qquad v_t =v.
$$

To proceed further,  we  need the following two equations for the variables $T_0\equiv vu$ and $ \theta\equiv v_x v^{-1}$:
\begin{equation}
{T_0}_x=\theta T_0-T_0 \rig{\theta}{},\qquad
\theta_y=T_0-\lef{T_0}{\:}.
\label{3}
\end{equation}
These equations are nothing but the  mapping (\ref{10}) rewritten in terms of
$T_0$ and $\theta$.

Now let us express ${\al_0}_y$ as
\begin{equation}
{\al_0}_y=\al_1\lef{T_0}{\:}+T_0\be_1.
\label{dec}
\end{equation}
  One can think of this substitution as  a decomposition of a vector in a two-dimensional space, where $\lef{T_0}{\:}$ and $T_0$ play the role of basis vectors.  Next, we substitute ${\al_0}_y$ in this form into the right hand side of
the second equation in~(\ref{2}), use  Eqs.~(\ref{3}) to express $\lef{T_0}{\:}_x $ and $T_{0x}$ in terms of $\theta$ and $T_0$, and equate to zero the coefficients multiplying $\lef{T}{}_0$ from the left and
  $T_0$ from the right in the resulting equation. The last step might not be a necessary condition, i.e. we can
potentially 
loose some solutions of the symmetry equations~(\ref{13}). Nevertheless, even with this restriction we obtain an infinite hierarchy  of nontrivial solutions of the symmetry equations~(\ref{13}). 

We have
\begin{equation}
\begin{array}{l}
{\al_1}_x=\al_0-\lef{\al}{}_0+\theta\al_1-\al_1\lef{\theta}{},\\

{\be_1}_x=\al_0-\rig{\al_0}{\!}+\rig{\theta}{}\be_1-\be_1\theta.
\end{array}\label{1000}
\end{equation}
Rewriting the second equation in (\ref{1000}) as 
$$
{\lef{\be}{}_1}_x=-(\al_0-\lef{\al}{}_0)+\theta\lef{\be}{}_1-
\lef{\be}{}_1\lef{\theta}{},
$$
we note that for $\lef{\be}{}_1=-\al_1$ the second equation in~(\ref{1000}) follows from the first. Looking for  solutions such that  $\lef{\be}{}_1=-\al_1$ and differentiating the first
equation in (\ref{1000}) with respect to   $y$, we find
\begin{equation}
\begin{array}{l}
\be_1=-\rig{\al_1}{\!},\\
{\al_1}_{xy}=(\al_1-\lef{\al}{}_1)\lef{T_0}{2}+T_0(\al_1-\rig{\al_1}{\!})+
\theta{\al_1}_y -{\al_1}_y
\lef{\theta}{}.\\
\end{array} \label{4}
\end{equation}
These equations have the same structure as Eqs.~(\ref{2}).  They also have an obvious solution  $\al_1=-\be_1=1$. 
  From this solution, using Eqs.~(\ref{dec}), (\ref{f}), and
the first equation in (\ref{2}), we find another solution of  the symmetry equations~(\ref{13}), i.e. the next term in our hierarchy of integrable systems
$$
u_t=F_1=u{\al_0}^{(1)}=u\int(\lef{T_0}{\:}-T_0)dy,\quad v_t=F_2={\be_0}^{(1)}v=\left[\int(\rig{T_0}{\:}-T_0)dy\right]v.
$$

Substituting
${\al_1}_y=\al_2\lef{T_0}{2}+T_0\be_2$ into  Eqs.~(\ref{4}) one can  continue to generate solutions of the symmetry equations~(\ref{13}) by the same
method that lead us from Eqs.~(\ref{2}) to (\ref{4}).  At the $n$th step we arrive at  
\begin{equation}
\begin{array}{l}
\be_n=-\rig{\al_n}{\!},\\
{\al_n}_{xy}=(\al_n-\lef{\al}{}_n)\lef{T_0}{(n+1)}+T_0(\al_n-\rig{\al_n}{\!})
+\theta{\al_n}_y-{\al_n}_{y}\lef{\theta}{n},\\
\end{array} \label{5}
\end{equation}
with a solution $\al_n=-\be_n=1$.
 This solution gives an expression for ${\al_0}^{(n)}$ that contains $2^n$ terms and can be written symbolically as
\begin{equation}
\begin{array}{rcl}
\displaystyle {\al_0}^{(n)}&=&\displaystyle (-1)^n \prod_{k=1}^n \left\{1-L_k
\exp\left[kd_k+\sum_{j=k+1}^n d_k\right]\right\}\times \\&&\\
&&\displaystyle \int dy \left(T_0 \int dy \left(\rig{T_0}{\!}\int dy \left(...\int dy
\left(\rig{T_0}{(n-1)}\right)...\right)\right)\right),
\label{20}
\end{array}
\end{equation}
where  the operator $\exp d_p$ denotes the application of the mapping (\ref{10}) to the function under the $p$th
integral, i.e.
$$
\dots\int dy\lef{T_0}{p} \dots \longrightarrow \dots\int dy\lef{T_0}{(p+1)}\dots,
$$
and the operator $L_r$  transposes  multipliers  inside the $r$th brackets,
$$
(A_1(\dots(A_r[\dots])\dots))  \longrightarrow (A_1(\dots([\dots]A_r)\dots)).
$$
The following multiplication rule applies:
$$
L_i \exp[...]_1 L_j\exp[...]_2=L_i L_j \exp\left[ [...]_1+[...]_2\right].
$$
  $L_n$ is an identity, $L_n=1$, since the $n$th bracket  contains a single operator $\rig{T_0}{(n-1)}$. Operators $\exp d_k$ mutually commute. For example,
$$
e^{d_1+d_2} e^{2 d_2}= e^{d_1} e^{d_2}  e^{d_2}  e^{d_2} = e^{d_1+3 d_2}.
$$

The comparison of Eq.~(\ref{20}) with  Eq.~(3.9) from \cite{14} shows that the only difference is that in the  noncommutative case we need to introduce  operators $L_i$ to maintain the proper ordering of the multipliers.

\subsection{Examples}
\label{DT_sec3}

Here we write down the first four  members in the hierarchy of integrable systems invariant with respect to the Darboux-Toda mapping~(\ref{10}). The first two ($n=0$ and $n=1$) are linear, while the rest ($n\ge 2$) are  not.

\subsubsection*{n=0}

$$
v_t=v,\qquad u_t=-u.
$$
\subsubsection*{n=1}
$$
v_t=v_x,\qquad u_t=u_x.
$$
\subsubsection*{n=2}
$$
v_t=v_{xx}-2\left[\int (vu)_x dy\right] v, \quad u_t=-u_{xx}+2u\int (vu)_x dy.
$$
These equations constitute the matrix Davey-Stewartson system~\cite{12}, a generalization of the well-known Davey-Stewartson 
system~\cite{12dss}
to the noncommutative case.

\subsubsection*{n=3}

$$
v_t=v_{xxx}-3\left[ \int (vu)_x dy\right] v_x-3\left[\int (v_x u)_x dy\right] v-
$$
$$
-3\left\{\int \left[vu\int (vu)_x dy-\left(\int (vu)_x dy\right) vu\right]dy\right\} v,
$$
\vspace{1em}
$$
u_t=-u_{xxx}-3u_x \int (vu)_x dy-3u\int (v_x u)_x dy-
$$
$$
-3u\int \left[vu\int dy (vu)_x-\left( \int dy (vu)_x \right) vu\right]dy.
$$
When $u$ and $v$ are scalar valued functions, these equations reduce to the Veselov-Novikov system.

\section{Darboux-Toda mapping in the two-dimensional superspace}
\label{super_sec}

Here we  analyze the situation, when noncommuting objects under
consideration in addition to the usual space-time coordinates $x, y$, and $t$
also depend on two Grassman variables $\theta_+$ and $\theta_-$.

\subsection{Definitions and the formulation of the problem}
\label{super_sec1}

 We  consider the
following mapping:
  \begin{equation}
  \begin{array}{cc}
\lef{u}{}= v^{-1},&\lef{v}{}=-\left[D_-\left(\left[D_+v\right] v^{-1}\right)+vu\right]v\equiv
v[D_+(v^{-1}D_-v)-uv],
  \end{array}
  \label{31}
  \end{equation}
where
$$
D_+=\frac{\partial}{\partial\theta_+}+\theta_+\frac{\partial}{\partial x},\quad
D_-=\frac{\partial}{\partial\theta_-}+\theta_-\frac{\partial}{\partial y},\quad
D^2_+=\frac{\partial}{\partial x},\quad D^2_-=\frac{\partial}{\partial y}.
$$
The rest of notations in this section is the same as in Section~\ref{DT_sec1}.  

The mapping inverse to the transformation (\ref{31}) is 
  \begin{equation}
  \begin{array}{cc}
\rig{v}{}= u^{-1},&\rig{u}{}=-[D_+([D_-u] u^{-1})+uv]u\equiv
u[D_-(u^{-1}D_+u)-vu],
  \end{array}
  \label{33}
  \end{equation}
while the symmetry equations to be solved now have the following form:
 \begin{equation}
\begin{array}{rcl}
\lef{F_1}{\:}&=
& -v^{-1}F_2 v^{-1},\\ 
\lef{F_2}{\,}&=&
F_2[D_+(v^{-1}D_-v)-uv]+v[D_+(-v^{-1}F_2v^{-1}D_-v)+\\&&+D_+(v^{-1}D_-F_2)-F_1v-
uF_2].
\end{array}   \label{35}
\end{equation}

\subsection{Solution of the symmetry equations}
\label{super_sec2}

Here we construct a hierarchy of integrable systems invariant under the mapping (\ref{31}).
The technique of solving the symmetry equations (\ref{35}) is essentially the same as   in the previous section.
However, there is an interesting difference --  solutions of these equations  now  obtain   only at even steps,
when unknown functions are bosonic. 
 
 Substituting $F_1=u\be_0$ and $F_2=\al_0 v$ into the symmetry equations (\ref{35}), we obtain [cf. Eqs.~(\ref{f})],
\begin{equation}
\begin{array}{l}
\be_0=-\rig{\al_0}{\!},\\ 
D_+D_-\al_0=(\lef{\al}{}_0-\al_0)\lef{T_0}{\:}+
T_0(\al_0-\rig{\al_0}{\!})+\theta D_-\al_0+D_-\al_0\theta, 
\label{62}
\end{array}
\end{equation}
where $T_0=vu$ and  $\theta=(D_+v) v^{-1}$. Eqs.~(\ref{62}) have a solution
$\al_0=-\be_0=1$ which yields $F_1=-u$ and $F_2=v$.

Mapping (\ref{31})  in terms of variables $T_0$ and $\theta$  reads [cf. Eqs.~(\ref{3})],
\begin{equation}
D_+T_0=\theta T_0-T_0 \rig{\theta}{},\qquad
D_-\theta=-T_0-\lef{T_0}{\:}.
\label{43}
\end{equation}

Next, we introduce new variables $\al_1$ and $\be_1$  through the equation
$D_-\al_0=\al_1\lef{T_0}{\:}+T_0\be_1$. Substituting this expression  for $D_-\al_0$ into the second equation in (\ref{62}), using relations (\ref{43}), and equating to zero the coefficients 
multiplying $T_0$ from the right  and $\lef{T_0}{\:} $ from the left,  we obtain
$$
D_+\al_1=\lef{\al}{}_0-\al_0+\theta\al_1+\al_1\lef{\theta}{},
$$
$$
D_+\be_1=\al_0-\rig{\al}{}_0+\rig{\theta}{}\be_1+\be_1\theta.
$$
For $\lef{\be}{}_1=\al_1$, the second of these relations  follows from the first. 
In this case, acting on the equation for $\al_1$ with $D_-$, we find
\begin{equation}
\begin{array}{l}
\be_1=\rig{\al_1}{\!},\\
-D_+D_-\al_1=(\al_1+\lef{\al}{}_1)\lef{T_0}{2}-T_0(\al_1+\rig{\al_1}{\!})+
D_-\al_1\lef{\theta}{}-\theta D_-\al_1.\\
\end{array} \label{44}
\end{equation}
These are  typical equations one obtains at odd steps of the procedure (step one in the present case). Comparing  them with Eqs.~(\ref{4}) of the previous section, we
see that Eqs.~(\ref{44}), in contrast to Eqs.~(\ref{4}), do not have
numerical solutions (since $\al_1$ and  $\be_1$ are fermionic functions).
Nevertheless, the procedure can be continued by writing 
$D_-\al_1=\al_2\lef{T_0}{2}+T_0\be_2$. We obtain
$$
-D_+\al_2=\lef{\al}{}_1+\al_1-\theta\al_2+\al_2\lef{\theta}{2},
$$
$$
-D_+\be_2=-(\al_1+\rig{\al_1}{\!})-\rig{\theta}{}\be_2+\be_2\lef{\theta}{}.
$$
For a solution such that $\lef{\be}{}_2=-\al_2$, we have
\begin{equation}
\begin{array}{l}
\be_2=-\rig{\al_2}{\!},\\
D_+D_-\al_2=(\lef{\al}{}_2-\al_2)\lef{T_0}{3}+T_0(\al_2-\rig{\al_2}{\!})+
D_-\al_1\lef{\theta}{2}+\theta D_-\al_2.\\
\end{array} \label{54}
\end{equation}
Eqs.~(\ref{54}) have a solution $\al_2=-\be_2=1$ that corresponds to the 
trivial solution of the symmetry equations~(\ref{35}): $F_1=au_x+bv_y, F_2=av_x+bv_y$ [see the discussion bellow Eq.~(\ref{13})].

Other equations obtained at even steps are similar to Eqs.~(\ref{54}) and have a solution
$\al_{2k}=-\be_{2k}=1$. For $k>1$ these solutions generate  nontrivial solutions of the symmetry equations~(\ref{35}) [see, e.g., the $k=2$ example in the next subsection].
In general, equations for $\al_{n}$ and $\be_{n}$ are [cf.~(\ref{5})]
 $$
\begin{array}{l}
\be_n=(-1)^{n+1}\rig{\al_n}{\!},\\
\\
D_+D_-\al_n=\left[(-1)^n\lef{\al}{}_n-\al_n\right]\lef{T_0}{n+1} +
T_0\left[\al_n+(-1)^{n+1}\rig{\al}{}_{n}\right]+\\
\\
\qquad\qquad\quad+(-1)^nD_+\al_n\lef{\theta}{n}+\theta D_-\al_n.\\
\end{array} 
$$
For $n=2k$ these equations have a solution $\al_{2k}=-\be_{2k}=1$, which gives the $k$th term of the hierarchy [cf.~(\ref{20})],
\begin{equation}
\begin{array}{rcl}
{\al_0}^{(k)}&=&(-1)^k \prod_{j=1}^{2k} \left\{1-(-1)^jL_j
\exp\left[jd_j+\sum_{j=k+1}^{2k} d_k\right]\right\}\times \\&&\\
&&\times D_-^{-1}\left(T_0 D_-^{-1}\left(\rig{T_0}{\!}D_-^{-1} \left(...D_-^{-1}
\left(\rig{T_0}{(2k-1)}\right)...\right)\right)\right).
\label{320}
\end{array}
\end{equation}
 The meaning of notations here is the same as in the formula~(\ref{20}).

\subsection{Examples}
\label{super_sec3}

Here we list the first three terms of the hierarchy of integrable systems invariant under the superspace version of the Darboux-Toda mapping (\ref{31}).

\subsubsection*{k=0}
\label{DT_ss3}

$$
v_t=v,\qquad u_t=-u.
$$

\subsubsection*{k=1}

$$
v_t=v_x,\qquad u_t=u_x.
$$

\subsubsection*{k=2}

$$
v_t=v_{xx}-2D_-^{-1}(vu)_x D_+v-2[D_-^{-1}(vD_+ u)] v+
$$
$$
+2D_-^{-1} \left\{vuD_-^{-1}(vu)_x +\left[D_-^{-1}(vu)_x \right] vu\right\},
$$
\vspace{1em}
$$
u_t=-u_{xx}+D_+uD_-^{-1}(vu)_x -2uD_-^{-1}(D_+vu)_x-
$$
$$
-2uD_-^{-1} \left\{vuD_-^{-1}(vu)_x +\left[D_-^{-1}(vu)_x \right] vu\right\}.
$$

\section{Conclusion}

The main  result of the present paper is the explicit construction
 of \textsl{quantum} integrable systems invariant with respect to the Darboux-Toda mappings~(\ref{10}) and (\ref{31}) [see Eqs.~(\ref{20}) and (\ref{320}) and examples in Subsections~\ref{DT_sec3} and \ref{DT_ss3} above].

 It is interesting that the technique we used to derive the hierarchies of integrable systems 
to a certain extent  resembles a computer algorithm -- there are many identical operations that can be
interrupted  at an arbitrary step. 
The structure of the group of integrable mappings, more precisely  of the
connection between an integrable system and its
symmetry equations, is encrypted in this ``algorithm''. Its translation   into the group-theoretic language would bring us closer to 
understanding  the role of integrable mappings in the theory of integrable systems.

It is well known that there is a close relation between quantum integrable systems and quantum algebras (see, e.g., \cite{16, 16_1}).  Therefore, it would be interesting to find the connection 
 between the approach of this paper and the well developed
formalism of quantum algebras.   

\section{Acknowledgments}
 The authors thank the Russian Foundation for Fundamental Research
for a support trough  Grant 95--01--00249.


\begin{thebibliography}{99}

\bibitem{1}
D. B. Fairlie and A. N. Leznov, The integrable mapping as the discrete group of inner symmetry of integrable systems, \href{https://doi.org/10.1016/0375-9601(95)00114-I}{  Phys. Lett. A,  \textbf{199}, 360 (1995)}.

\bibitem{1_1}

A. N. Leznov, The new look on the theory of integrable systems, \href{https://doi.org/10.1016/0167-2789(95)00152-T}{Physica D \textbf{87}, 48 (1995)}.
 

\bibitem{10}
 A. N. Leznov and E. A. Yuzbashyan, The general solution of two-dimensional matrix Toda chain equations with fixed ends, \href{https://link.springer.com/article/10.1007/BF00750841}{Lett. in Math. Phys. \textbf{35}, 345, (1995)}.

\bibitem{12}
A. N. Leznov and E.A. Yuzbashyan,   Multi-soliton solutions of the two-dimensional matrix Davey-Stewartson equation,
\href{https://doi.org/10.1016/S0550-3213(97)00264-2}{Nucl. Phys. B \textbf{496}, 643 (1997)}; \href{ 	
https://doi.org/10.48550/arXiv.hep-th/9612107}{arXiv:hep-th/9612107}.

\bibitem{12dss}

A. Davey and K. Stewartson, On three-dimensional packets of surface waves, \href{https://doi.org/10.1098/rspa.1974.0076}{Proc. Roy. Soc. A \textbf{338}, 101
(1974)}.


\bibitem{14}
A. N. Leznov and E. A. Yuzbashyan,  Two-dimensional integrable Darboux-Toda substitution and Davey-Stewartson hierarchy of integrable systems,  \href{https://lss.fnal.gov/archive/other/ifve-95-28.pdf}{Preprint IHEP-95-28, Protvino 1995}.

\bibitem{16}
V. G. Drinfeld ``Quantum Groups'' In: Proc. of the Intern. Congress of
Mathematicians, Vol. 1, p. 798, New York 1986.

\bibitem{16_1}

L. D. Fadeev, N. Yu. Reshetikhin, and L. A. Takhtajan, Leningrad Math. J. \textbf{1}, 193 (1990).

\end{thebibliography}
\end{document}